\documentclass[11pt]{article}
\usepackage{moriond2000,epsfig}

\def\be{\begin{equation}}
\def\ee{\end{equation}}
\def\bea{\begin{eqnarray}}
\def\eea{\end{eqnarray}}

\begin{document}
\vspace*{4cm}
\title{Microlensing of Quasars}

\author{ Joachim Wambsganss}

\address{Universit\"at Potsdam, Institut f\"ur Physik, Am Neuen
Palais 10, 14469 Potsdam\\
and \\
	Max-Planck-Institut f\"ur Gravitationsphysik, 
	Am M\"uhlenberg 1,
	14476 Golm,
	Germany}

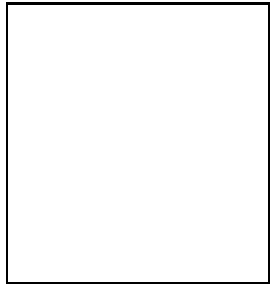
\maketitle

\abstracts{There are two possible causes of 
variability in gravitationally lensed quasars: 
intrinsic fluctuations of the quasar and ``microlensing"
by compact objects along the line of sight. 
If disentangled from each other, microlens-induced
variability can be used to study two cosmological issues of
great interest,  
the size and brightness profile of quasars on one hand,
and the distribution of compact (dark)  matter along the line of sight.
Here we present a summary of recent 
observational evidence for quasar microlensing
as well as 
of theoretical progress in the field.
Particular emphasis is given to the questions which
microlensing can address  regarding the search for  dark
matter, both in the halos of lensing galaxies and in a
cosmologically distributed form.
A discussion
of desired observations and required theoretical studies
is presented as a conclusion/outlook.
}

\section{What is Microlensing of Quasars?}

\subsection{Mass, length and time scales}
The lensing effects on  quasars by
compact objects in the mass range 
$  10^{-6} \le m/M_{\odot} \le 10^3$
is usually called ``quasar microlensing".
The microlenses can be ordinary stars, brown dwarfs, 
planets, black holes,
molecular clouds, globular clusters 
or other compact mass concentrations (as long as their
physical size is smaller than their Einstein radius).
In most practical cases, the microlenses are part of a galaxy which 
acts as the main (macro-)lens.  However, microlenses could also
be located in, say, clusters of galaxies or they could even be 
imagined ``free floating" and filling intergalactic space.

The relevant length scale for microlensing (in the quasar plane)
is the Einstein radius of the lens: 
	$$ r_E = 
\sqrt{ { {4 G M } \over {c^2} } { {D_S D_{LS} \over D_L}  } }
\approx 
	4 \times 10^{16} \sqrt{M / M_\odot} \rm \,  cm, $$
where ``typical" lens and source redshifts of $z_L \approx 0.5$ 
and
$z_S \approx 2.0$  are assumed for the numerical value
on the right hand side ($G$ and $c$ are the gravitational
constant and the velocity of light, respectively; $M$ is the mass of
the lens, $D_L$, $D_S$, and $D_{LS}$ are the angular diameter distances
between observer -- lens, observer -- source, 
and lens -- source, respectively).
Quasar microlensing turns out to be an interesting phenomenon, 
because (at least) the size of the
continuum emitting region of quasars is 
comparable to or smaller than the Einstein radius of 
stellar mass objects.

This length scale translates into an angular scale of
	$$ \theta_E = r_E/D_S  
		\approx  10^{-6} \sqrt {M /M_\odot} \ \ \rm arcsec. $$
It is obvious that image splittings on such angular
scales  cannot be observed directly. What makes microlensing observable
anyway is the fact that observer, lens(es) and source move relative to each
other. Due to this relative motion, the micro-image configuration 
changes with time, and so does the total magnification, i.e. the
sum  of the magnifications of all the micro-images. This change
in magnification over time can be measured:
microlensing is a ``dynamical" phenomenon. 

There are two time scales involved: the standard lensing
time scale $t_E$ is the time it takes the source to cross the
Einstein radius of the lens, i.e. 

	$$ t_E = r_E/v_{\perp, \rm eff}   \approx  15 \sqrt {M / M_\odot}  v_{600}^{-1} \ \ \rm  years, $$
where the same assumptions are made as above, 
and the effective relative transverse velocity 
$v_{\perp, \rm eff}$ is parametrized in 
units of 600 km/sec: $v_{600}$.
This time scale $t_E$ results in discouragingly large values. 
However, in practice 
we can expect fluctations on much shorter time intervals.
The reason is that the sharp caustic lines separate regions of low 
and high magnification.  Hence, 
if a source crosses such a caustic line, we can observe a large
change in magnification during the time $t_{cross}$
it takes the source to cross its own diameter $R_{source}$: 

	$$ t_{cross} 
= R_{source}/v_{\perp, \rm eff}   \approx   4 R_{15}  v_{600}^{-1} \ \ \rm  months.$$
Here the quasar size $R_{15}$ is parametrized in units of
$10^{15}$cm.

\subsection{Early Promises of Quasar Microlensing}

The early papers exploring microlensing  
made four predictions concerning the potential
scientific results. 
Microlensing should  help us to determine:
	1) the existence and effects of compact objects between 
		the observer and the source,
	2) the size of quasars,
	3) the two-dimensional brightness profile of quasars,
	4) the mass (and mass distribution) of lensing objects.
In Section 3 the observational results to date will be discussed in 
some detail.  It can be stated here that 1) has been achieved. 
Some limits on the size of quasars have been obtained, so 2) is partly
fulfilled. We are still (far) away from solving promise 3), and 
concerning point 4) it is fair to say that 
the observational results are consistent with certain
(conservative) mass ranges.

\subsection{Quasar Microlensing versus ``Local Group" Microlensing }

In most cases of quasar microlensing,
the surface mass density (or optical depth)
is of order unity. 
In contrast to that, 
the ``local group" microlensing deals with very low optical  depths,
where the action is due to single lenses or physical
binaries. 
Since there are interesting similarities as well 
differences between these two kinds of microlensing, 
in Table \ref{table1}  a few 
quantities relevant to microlensing are compared to each other
for the two regimes.
\begin{table}[thb]
\begin{center}
\begin{tabular}{|c||c|c|}
\hline\hline
&&\\
Lensing galaxy:  & Milky Way & Lens in Q0957+561 \\
&&\\
\hline\hline
&&\\
distance to Macho known? & no & yes \\
&&\\
velocity of Macho known? & no & (no) \\
&&\\
 mass?   &  ???   &  ???   \\
&&\\
 optical depth?  & $ \approx 10^{-6}$  & $\approx$ 1  \\
&&\\
Einstein angle (1 M$_\odot$)? & $\approx$ 1 milliarcsec  & $\approx$ 1 microarcsec  \\
&&\\
time scale?  & hours to years  & weeks to decades \\
&&\\
event? & individual/simple & coherent/complicated \\
&&\\
default light curve?      & smooth & sharp caustic crossing \\
&&\\
when/who proposed? & Paczy\'nski 1986 & Gott 1981 \\
&&\\
first detection?   & EROS/MACHO/OGLE & Irwin et al. 1989 \\
		   &                 1993 & \\
&&\\
\hline\hline
\end{tabular}
\caption{A few lensing properties for the
two regimes of microlensing 
are compared to each other: 
local group microlensing and quasar microlensing.}
\label{table1}
\end{center}
\end{table}

\section{Theoretical Work on Quasar Microlensing}
For a multiply imaged quasar, the surface mass density (or
``optical depth") at the position of an image is of order unity. If this
matter is made of compact objects in the range described above, 
microlensing is expected to be going on basically ``all the time",
due to the relative motion of source, lens(es) and observer. In addition,
this means that the lens action is due to a coherent effect 
of many microlenses,
because the action of two or more point lenses whose projected distances
are of order their Einstein radii combines in a very non-linear way
(cf. Wambsganss 1998). 

The lens action of more than two point lenses cannot be easily treated
analytically any more. Hence numerical techniques were developed in order
to simulate the gravitational lens effect of many compact objects.
Paczy\'nski (1986) had used a method to look for the extrema in the
time delay surface. Kayser, Refsdal, Stabell (1986), Schneider 
\& Weiss (1987) and Wambsganss (1990) had developed and applied
an inverse ray-shooting technique that
produced a two-dimensional magnification distribution in the source plane.
An alternative technique was developed by Witt (1993) and 
Lewis et al. (1993);
they solved the lens equation along a linear source track.
All the recent theoretical work on microlensing is based on either
of these techniques.

Recently,
Fluke \& Webster (1999) explored analytically  caustic crossing events for
a quasar. Lewis et al. (1998) showed that 
spectroscopic monitoring of multiple quasars can be used to probe the
broad line regions.
Wyithe et al. (2000a, 2000b) investigated and found 
limits on the quasar size and on the mass function  in Q2237+0305.

Agol \& Krolik (1999)
and 
Mineshige \& Yonehara (1999) developed techniques to recover
the one-dimensional brightness profile of a quasar, based on the
earlier work by Grieger et al. (1988, 1991). Agol \& Krolik showed
that frequent monitoring  of a caustic crossing event in many
wave bands (they used of order 40 simulated
data points in eleven filters over
the whole electromagnetic range), one can recover 
a map of the  frequency-dependent brightness distribution
of a quasar.
Yonehara (1999) in a similar approach explored the effect of 
microlensing on two different accretion disk models.
In another paper, Yonehara  et al. (1998)
showed that monitoring  a microlensing event in the 
X-ray regime can reveal structure of the quasar accretion
disk as small as AU-size.

\section{Observational Evidence for  Quasar Microlensing}
Fluctuations in the brightness of a quasar can have two causes: they
can be intrinsic to the quasar, or they can be microlens-induced.
For a single quasar image, the difference is hard to tell. However,
once there are two or more gravitationally lensed (macro-)images of a 
quasar, we have a relatively good handle to distinguish the two
possible causes of variability: any fluctuations due to
intrinsic variability of the quasar show up in all
the quasar images, after a certain time delay. 
(This argument could even be turned around: the measured time delays
in multiple quasars are the ultimate proof of the intrinsic
variability of quasars.)  
So once a time delay is measured in a multiply-imaged quasar system, 
one can shift the lightcurves of the different
quasar images relative to each other by the time delay, 
correct for 
the different (macro-)magnification, and subtract them from each
other. 
All remaining incoherent fluctuations 
in the ``difference lightcurve" 
can be contributed to microlensing.
In a few  quadruple lens systems we can detect microlensing
even without measuring the time delay:
in some cases the image arrangement is so symmetrical 
around the lens that any
possible lens model predicts very short time delays (of order days or
shorter), so that fluctuations in individual images that last 
longer than a day or so and are not followed by corresponding
fluctuations in the other images, can be safely attributed to microlensing.
This is in fact the case for the quadruple system Q2237+0305.

\subsection{The Einstein Cross: Quadruple Quasar Q2237+0305}
In 1989 the first evidence for quasar microlensing was
found by Irwin et al. (1989) in the quadruple quasar Q2237+0305:
one of the components showed fluctuations. In the mean time,
Q2237+0305 has been monitored by many groups
(Corrigan et al. 1991;  {\O}stensen et al. 1996;
Lewis et al. 1998).
The most recent (and  most exciting) results (Wozniak et al. 2000)
show that
all four images vary dramatically, going up and down like
a rollercoaster in the last three years: 
$ \Delta m_A \approx$ 0.6 mag, 
$ \Delta m_B \approx$ 0.4 mag, 
$ \Delta m_C \approx$ 1.3 mag (and rising?), 
$ \Delta m_D \approx$ 0.6 mag.

\subsection{The Double Quasar Q0957+561 }
The microlensing results for the double quasar Q0957+561 are not
as exciting. In the first few years 
there appears to
be an almost  linear change in the (time-shifted) brightness
ratio between the two images 
($ \Delta m_{AB} \approx 0.25$ mag over 5 years).
But since about 1991, this ratio
stayed more or less ``constant" within about 0.05 mag, so 
not much microlensing was going on in this system recently
(Schild 1996; Pelt et al. 1998; Schmidt \& Wambsganss 1998).
The possibility for some small amplitude rapid microlensing 
(cf. Colley \& Schild 2000)
cannot be excluded; however, one needs a very well determined time delay
and very accurate photometry, in order to confirm it. 

With numerical simulations and limits obtained from data of
three years of Apache Point monitoring data of Q0957+561, 
and based on the Schmidt \& Wambsganss (1998) analysis, 
Wambsganss et al. (2000)
extend the limits on the masses of 
``Machos" in the (halo of the) lensing
galaxy in 0957+561: the small ``difference" between the time-shifted 
and magnitude-corrected lightcurves of images A and B 
excludes a halo of the lensing galaxy made of compact objects with
masses of $10^{-7} M_\odot - 10^{-2} M_\odot$. Similar results were
found by Refsdal et al. (2000) based on an independent data set.

\subsection{Other multiple quasars/radio microlensing?}
A number of other multiple 
quasar systems are being monitored more or less
regularly. For some of them microlensing has been suggested (e.g.
H1413+117,  {\O}stensen et al. 1997;  or B0218+357, Jackson et al. 2000).
In particular the possiblity for ``radio"-microlensing appears
very interesting (B1600+434,  Koopmans~\&~de~Bruyn~2000),
because this is unexpected, 
due to the presumably larger source size of the radio 
emission region. The possibility of relativistic motion of radio
jets may make up for this ``disadvantage".

\section{Unconventional Quasar Microlensing}

\subsection{Microlensing in individual quasars? }
There were a number of papers interpreting the 
variability of individual quasars  as microlensing
(e.g., Hawkins \& Taylor 1997, Hawkins 1998). Although this
is an exciting possibility and it could help us detect a 
population of cosmologically distributed lenses, it is not
entirely clear at this point whether the observed fluctuations
can be really attributed to microlensing. After all, quasars are
intrinsically variable,
and the expected microlensing in
single 
quasars must be smaller than in multiply imaged ones, due
to the lower surface mass density. 
More studies are necessary to clarify this issue.

\subsection{``Astrometric Microlensing": Centroid shifts  }
An interesting aspect of microlensing was explored by 
Lewis \& Ibata (1998). They looked 
at centroid shifts of quasar images due to microlensing. At each 
caustic crossing, a new very bright image pair emerges or disappears,
giving rise to sudden changes in the ``center of light" positions.
The amplitude could be of order 100 microarcseconds or larger,
which should be observable with the 
SIM satellite (Space Interferometry Mission), to be launched
in June 2006.

\section{Quasar Microlensing: Now and Forever? }
Monitoring observations of various multiple 
quasar systems in the last decade have clearly established 
that the phenomenon of quasar microlensing exists. 
There are uncorrelated variations in multiple quasar
systems with amplitudes of more than a magnitude and time scales
of weeks to months to years. However, in order to get closer to a
really quantitative understanding, much better monitoring programs need
to be performed.

On the theoretical side, there are two important questions:
what do the lightcurves tell us about the lensing objects, and
what can we learn about the size and structure of the
quasar. As response to the first question,  
    numerical simulations are able to give a qualitative understanding
of the measured lightcurves (detections and non-detections), in general
consistent with ``conservative" assumptions about the object masses
and velocities. But
due to the large number of parameters (quasar size, masses of lensing
objects, transverse velocity) and due to the large variety of
lightcurve shapes, 
no satisfactory quantitative 
explanation or even prediction could be achieved. 
So far mostly ``limits" on certain parameters have been obtained.
The prospects of getting
much better lightcurves of multiple quasars, as shown by the OGLE 
collaboration, should be motivation enough to explore this 
direction in much more quantitative detail. 

The question of the structure of quasars deserves 
more attention. Here
gravitational lensing is in the unique situation to be able to explore
an astrophysical field that is unattainable by any other means. Hence
more effort should be put into attacking this problem. 
This involves much more ambitious observing programs, with the goal
to monitor caustic crossing events in many filters over the whole
electromagnetic spectrum, and to further develop numerical techniques
to obtain useful values for quasar sizes
and luminosity profiles
from unevenly sampled data in (not enough) different filters.


\end{document}